\documentclass[prb,twocolumn,showpacs,amsmath,amssymb,floatfix]{revtex4}

\usepackage{graphicx}
\usepackage{hyperref}

\hypersetup{%
  a4paper = {true},
  bookmarksnumbered = {true}, 
  bookmarksopen = {true}, 
  bookmarksopenlevel = {2}, 
  breaklinks = {true}, 
  colorlinks = {true}, 
  pdfauthor = {\textcopyright\ Carsten Raas and Götz S. Uhrig},
  pdfcreator = {\LaTeX\ with package \flqq hyperref\frqq},
  pdfkeywords = {DMRG, density-matrix renormalization, density matrix renormalization,
    D-DMRG, dynamic density-matrix renormalization, dynamic density matrix renormalization,
    correction vector, dynamics, spectral densities,
    SIAM, single-impurity Anderson model, single impurity Anderson model}
  pdfmenubar = {true},
  pdfstartpage = {1},
  pdfstartview = {FitH}, 
  pdfsubject = {Spectral Densities from Dynamic Density-Matrix Renormalization},
  pdftitle = {Spectral Densities from Dynamic Density-Matrix Renormalization},
  pdftoolbar = {true},
  plainpages = {false}, 
  urlcolor = {blue} 
}

\begin{document}

\title{\bf Spectral Densities from Dynamic Density-Matrix Renormalization}

\author{Carsten Raas}
\email[\emph{Electronic address: }]{raas@lusi.uni-sb.de}

\author{G\"otz S.~Uhrig}
\email[\emph{Electronic address: }]{uhrig@lusi.uni-sb.de}
\homepage[\\\emph{Homepage: }]{http://www.uni-saarland.de/fak7/uhrig/}

\affiliation{Institut f\"ur Theoretische Physik,
  Universit\"at zu K\"oln,
  Z\"ulpicher Str.\ 77,
  50937 K\"oln,
  Germany}
\altaffiliation[\emph{Present address: }]{Theoretische Physik FR 7.1,
  Geb\"aude 38,
  Universit\"at des Saarlandes,
  66123 Saarbr\"ucken,
  Germany}

\date{July 6, 2005}

\begin{abstract}
  Dynamic density-matrix renormalization provides valuable numerical
  information on dynamic correlations by computing convolutions of the
  corresponding spectral densities. Here we discuss and illustrate how and to
  which extent such data can be deconvolved to retrieve the wanted spectral
  densities. We advocate a nonlinear deconvolution scheme which minimizes the
  bias in the ansatz for the spectral density. The procedure is illustrated
  for the line shape and width of the Kondo peak (low energy feature) and for
  the line shape of the Hubbard satellites (high energy feature) of the single
  impurity Anderson model. It is found that the Hubbard satellites are strongly
  asymmetric.
\end{abstract}

\pacs{71.55.Ak, 71.27+a, 78.67.Hc, and 78.20.Bh} 


\maketitle


\section{Introduction}

The computation of spectral properties is a central issue in theoretical
physics. Many spectroscopic probes provide experimental information about the
investigated systems. In order to understand the meaning of such data it is
indispensable to be able to compute the corresponding quantities theoretically.
This task is particularly demanding if the system under study is characterized
by strong correlations. Then standard approaches like diagrammatic perturbation
theory have difficulties to provide quantitative results.

An archetypal class of strongly correlated systems are impurity models where a
small subsystem, the impurity, is coupled to a bath of degrees of freedom. The
discrete levels of the impurity are broadened due to the interaction with the
bath. The most fundamental fermionic representative of this class of models is
the single impurity Anderson model (SIAM) \cite{hewso93} where the impurity is
characterized by a single fermion level which can be empty, or singly occupied
with spin up or down, or doubly occupied. The energy of the doubly occupied
state is increased by the interaction energy $U$. The bath is a bath of
non-interacting fermions of either spin direction to which the impurity is
coupled by hybridization. It is the aim of this article to present an algorithm
to obtain the dynamics of a SIAM with high resolution on all energy scales.

The SIAM describes a plethora of physical problems. Historically it was used
for diluted magnetic impurities in metals, see e.g.\ Ref.\ 
\onlinecite{hewso93}. But it also describes the electronic transmission through
quantum dots, see e.g.\ Ref.\ \onlinecite{pusti04}. The smallness of the
quantum dot implies a small capacitance, hence a large charging energy which
represents the interaction energy $U$. The bath is given by the external leads.
The dynamic mean-field theory (DMFT) \cite{prusc95,georg96} represents
another broad and very active field where the SIAM occurs. In this approach,
as in all mean-field approaches, the lattice problem of strongly interacting
fermions is mapped onto an effective single-site problem, namely a SIAM. This
SIAM is linked to the original lattice problem by a self-consistency condition.
The clue is that the mean-field, the Green function of the bath, is a dynamic
quantity depending on frequency.

The above examples illustrate that it is very important to be able to compute
the dynamics of a SIAM in a reliable fashion. There are several numerical
approaches which can be applied. Among the most prominent ones are quantum
Monte Carlo (QMC) \cite{hirsc86} and the numerical renormalization group (NRG)
\cite{sakai89,costi90}. Both approaches are powerful but do not have a high
resolution away from the Fermi level. For QMC this is so since the dynamics is
computed in imaginary time and the analytic continuation to real frequencies
represents an ill-conditioned problem. Moreover, care must be taken to reach
low temperatures. The NRG can be used directly at zero temperature. But it is
set up to focus on the limit $\omega\to 0$. The energy levels kept are
broadened by a broadening which is proportional to the frequency which implies
that features at higher energies tend to be smeared out \cite{raas04a}.

We investigate here a third complementary numerical approach given by the
dynamic density-matrix renormalization (D-DMRG)
\cite{hallb95,ramas97,kuhne99a,hovel00,jecke02,raas04a,nishi04a}. Note that
there is no complete consensus on the nomenclature. Jeckelmann uses the term
`dynamic density-matrix renormalization group' only for the approach using a
variational principle \cite{jecke02} while the approach used here is called
correction vector density-matrix renormalization. We consider, however,
`dynamic density-matrix renormalization group' to be the general term for all
algorithms computing dynamic correlations by DMRG in the frequency domain
\cite{hallb95,ramas97,kuhne99a,jecke02} as opposed to DMRG approaches computing
the dynamics in real time \cite{white04a,daley04} which efficiently
implement ideas of Vidal \cite{vidal03,vidal04}. The approach used in this
article determines the dynamics at zero temperature by computing the
expectation values in the local propagator. This can be realized by targeting
not only at the ground state and the excited state, but also at the resolvent
applied to the excited state. This additional targeted state is called the
correction vector.

The main limitation of the D-DMRG is that one cannot obtain data for purely
real frequencies but only for frequencies with a certain imaginary part. Hence
the extraction of the behavior at purely real frequencies is one of the main
problems to be solved in using the D-DMRG \cite{gebha03,raas04a,nishi04a}. It
is the main aim of the present paper to discuss and to compare various
algorithms to achieve this extraction. In particular, we will present a
nonlinear approach from the family of maximum entropy methods. This approach
provides a \emph{continuous}, \emph{positive} ansatz for the wanted spectral
density with the \emph{least bias} (LB).

In the following section \ref{sec:model} we will present the model and we will
discuss the observable of interest. Next in Sect.\ \ref{sec:extract}, we will
discuss various extraction schemes, linear and non-linear ones. The features of
these schemes will be illustrated by some toy spectral densities for which the
broadened and the unbroadened data is analytically available. The results by
the LB algorithm for the Kondo peak and the Hubbard satellites will be
presented in Sects.\ \ref{sec:kondo} and \ref{sec:satellit}, respectively.
Finally, a short summary will be given in Sect.\ \ref{sec:summary}.


\section{Model and Observable}
\label{sec:model}

We want to illustrate the above general remarks by D-DMRG calculations and the
subsequent data analysis. As motivated in the introduction the single-impurity
Anderson model (SIAM) at half-filling is a very good and interesting testing
ground. We will focus on
\begin{align}
  \nonumber
  \mathcal{H} = & U
  \left(n_{d,\downarrow}-\frac{1}{2}\right)
  \left(n_{d,\uparrow}-\frac{1}{2}\right)
  + V
  \sum_\sigma\left(
    d_{\sigma}^{\dagger} c_{0,\sigma}^{\phantom{\dagger}} + \text{h.c.}
  \right)\\
  & + \sum_{n=0,\sigma}^\infty \gamma_{n+1}\left(
    c_{n,\sigma}^\dag c_{n+1,\sigma}^{\phantom{\dagger}} + \text{h.c.}
  \right)
  \label{eq:hamilton}
\end{align}
with arbitrary symmetric density of states (DOS) $\rho_0(\omega)$ of the free
($U=0$) one-particle Green function $G_0(\omega)$ of the $d$-electron. The
$d$-electron represents the impurity which is correlated due to the interaction
$U>0$. The bath is represented by the coefficients $\gamma_n\ge0$ in
(\ref{eq:hamilton}). They are the coefficients of the continued fraction of the
hybridization function $\Gamma(\omega)$ (cf.\ Ref.\ \onlinecite{raas04a}). Any
hybridization function with symmetric imaginary part $\rho_0(\omega) :=
-\pi^{-1}\text{Im}\, G_0(\omega+i0+)$ can be represented by an appropriate
choice of the $\gamma_n$. Hence the representation of the bath as semi-infinite
chain does not restrict the generality of the model.

The dynamics we wish to compute is the dynamics of the fermionic
single-particle propagator of the $d$-electron. Aiming at the properties at
$T=0$ it reads
\begin{align}
  \label{eq:greendef}
  \begin{split}
    G(\omega+i\eta) &= \left\langle 0\left|
        d_\sigma \frac{1}{\omega+i\eta-(\mathcal{H}-E_0)}
        d^\dagger_\sigma \right|0 \right\rangle \\
    &+ \left\langle 0\left|
        d^\dagger_\sigma \frac{1}{\omega+i\eta+(\mathcal{H}-E_0)}d_\sigma
      \right|0\right\rangle\ .
  \end{split}
\end{align}
Here the ground state is denoted by $|0\rangle$ and its energy by $E_0$. Since
we focus at a spin-disordered solution the propagator has no dependence on the
spin index $\sigma$. Hence, it is not denoted as argument of $G$. 
The frequencies $\omega$ and $\eta$ are real. 
The standard retarded Green function is obtained
for $\eta\to0+$
\begin{equation}
  G_\text{R}(\omega) = \lim_{\eta\to0+}G(\omega+i\eta)\ .
\end{equation}
The quantity we are looking for is the spectral density $\rho(\omega):=
-\pi^{-1}\text{Im}\,G_\text{R}(\omega)$. If necessary the real part can be
obtained from the Kramers-Kronig relation.

The D-DMRG provides data points at given values of $\omega=\xi_i$ for finite
values of $\eta_i$
\begin{equation}
  \label{eq:convolv0}
  g_i = -\frac{1}{\pi}\text{Im}\, G(\xi_i+i\eta_i) = \frac{1}{\pi}\
  \int_{-\infty}^\infty 
  \frac{\eta_i \rho(\omega)\text{d}\omega}{(\xi_i-\omega)^2+\eta_i^2}\ ,
\end{equation}
where we used the Hilbert representation in the second equation. No data can be
obtained directly at $\eta=0$ since the inversion of the Hamiltonian is
singular and cannot be achieved numerically in a stable way. Henceforth, we
will call data at finite values of $\eta$ \emph{raw} data. One way to extract
the physically relevant data on the real axis is to look at a sequence of
decreasing values of $\eta$ in order to extrapolate the result \cite{nishi04a}
to $\eta=0$. This approach, however, is time-consuming and requires many
resources, in particular, if one is interested in the whole spectral density.
So the line followed in the present article is to use the raw data as input of
a generalized scheme to extract the information on the spectral density
$\rho(\omega)$.


\section{Extraction Schemes}
\label{sec:extract}

We consider two classes of extraction schemes, linear ones and non-linear ones.
Linearity means that there is a linear relation between the raw data and the
wanted spectral density.


\subsection{Linear Extraction Schemes}


\subsubsection{Deconvolution}

If the unavoidable imaginary part $\eta$ is constant for all frequencies Eq.\ 
(\ref{eq:convolv0}) becomes
\begin{equation}
  \label{eq:convolv}
  g_i = -\frac{1}{\pi}\text{Im}\, G(\xi_i+i\eta) = \frac{1}{\pi}\
  \int_{-\infty}^\infty 
  \frac{\eta \rho(\omega)\text{d}\omega}{(\xi_i-\omega)^2+\eta^2}
\end{equation}
so that the raw data is the convolution of the true spectral density
$\rho(\omega)$ with the Lorentzian
\begin{equation} 
  L_\eta(\omega) = \frac{1}{\pi}\frac{\eta}{\omega^2+\eta^2}
\end{equation}
of width $\eta$. Hence the necessary step for retrieving $\rho(\omega)$ is a
deconvolution. It can be achieved in various ways. One standard way is to
deconvolve the raw data. This is done in the time domain reached by Fourier
transform because the convolution in Eq.\ (\ref{eq:convolv}) becomes a product
in the time domain
\begin{equation}
\label{timedomain}
\rho(\tau)_\text{raw} = \exp(-\eta|\tau|) \rho(\tau)\ ,
\end{equation}
where we use $\rho(\tau)$ for the Fourier transform of $\rho(\omega)$ and
$\rho(\tau)_\text{raw}$ for the Fourier transform of the raw data. The raw data
$\{g_i\}$ is obtained in the first place as a discrete set. In order to obtain
a quasi continuous distribution we interpolate the discrete set $\{g_i\}$ by
splines \cite{press92} which leads to $\rho(\omega)_\text{raw}$. The Fourier
transforms are most efficiently done by Fast Fourier algorithms \cite{press92}.
The actual deconvolution is done by dividing by $\exp(-\eta|\tau|)$ which
inverts Eq.\ (\ref{timedomain}). Then one transforms back to the frequency
domain. So this procedure is very efficient and straightforward \cite{raas04a}.

The restriction to be kept in mind is that splining and deconvolution cannot
create information where no information was present before. If the input data
is not precise enough or if the spline does not approximate the true continuous
function $\rho(\omega)_\text{raw}$ well enough the deconvolution will fail to
produce reasonable results. In practice this is seen in unreasonable values of
$\rho(\tau)$ after the division by $\exp(-\eta|\tau|)$ because this division
amplifies any inaccuracy for large values of $|\tau|$. This problem is
circumvented by a suitable low-pass filter $p_{\tau_0,\Delta\tau}(\tau)$
which suppresses inaccurate values
\begin{equation}
  \rho(\tau)_\text{raw} \exp(\eta|\tau|)\to
  \rho(\tau)_\text{raw} \exp(\eta|\tau|) p_{\tau_0,\Delta\tau}(\tau) 
\end{equation}
at large values of $|\tau|$ beyond $\tau_0$ on the scale $\Delta \tau$. Of
course, this implies that only a certain resolution in frequency can be
achieved. In view of the inevitable inaccuracies of any numerical calculation
one has to accept such a bound to the enhancement of the resolution.
Nevertheless, the deconvolution enhances the resolution considerably and the
final curve obtained is continuous for all practical purposes since the
interpolation allows to make the grid as fine as needed.

\begin{figure}[t]
  \begin{center}
    \includegraphics[width=\columnwidth,angle=0]{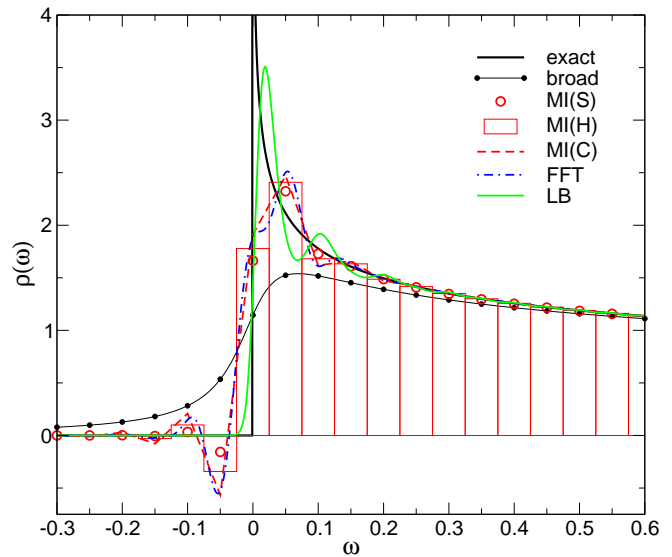}
    \caption{\label{fig:decon1} 
      Doniach-\v{S}unji\'c line shape (Eq.\ (\ref{eq:doniach})) for
      $\alpha=1/4$ with (thin black solid line with circles representing the
      raw data at $\xi_i=0.05i$) and without broadening 
      (thick black solid line)
      $\gamma=\eta=0.05$. The results of various schemes to retrieve the
      unbroadened line are shown: by deconvolution via fast Fourier transform
      (FFT), by matrix inversion assuming spikes (MI(S)) or histograms (MI(H))
      or a piecewise linear continuous function (MI(C)), by the non-linear
      least-bias algorithm (LB). All schemes are described in the main text.}
  \end{center}
\end{figure}
The procedure is illustrated in Fig.\ \ref{fig:decon1} where we display a power
law singularity $s(\omega) = \omega^{-\alpha}\Theta(\omega)$ exactly and
convolved by a Lorentzian $L_\gamma(\omega)$ of width $\gamma=0.05$
\begin{equation}
  \label{eq:doniach}
  s(\omega;\gamma) = \frac{
    \cos\left[
      \pi(1-\alpha)/2-\alpha\arctan(\omega/\gamma)
    \right]}
  {\sin\left(\pi\alpha\right)(\omega^2+\gamma^2)^{\alpha/2}}\ . 
\end{equation}
This line shape is well known in photoelectron spectroscopy \cite{hufne03},
named the Doniach-\v{S}unji\'c line shape \cite{donia70}. Note that we changed
the normalization so that (\ref{eq:doniach}) represents exactly the convolution
of $\omega^{-\alpha}\Theta(\omega)$ with $L_\gamma(\omega)$. The raw data
$g_i=s(\xi_i;\gamma=\eta)$ that we use for these curves is obtained
analytically at $\xi_i = i*0.05$ in the interval $\xi_i\in [-3,5]$, see the
curve with the circle symbols. We do not use real D-DMRG here since we want
first to illustrate the extraction schemes under ideal circumstances. The
effect of inaccuracies will be discussed below.

In judging the effect of the deconvolution one must keep in mind that the
reconstruction of a diverging singular line is the worst case for any
algorithm. We have chosen the Doniach-\v{S}unji\'c line shape for illustration
in order to highlight the differences in the various schemes. Below (Fig.\ 
\ref{fig:decon2}) we will present results also for a smoother curve to show
that such a curve can be reconstructed in a quantitatively reliable way.

The dashed-dotted line is the result of the above described deconvolution using
the low-pass filter
\begin{equation}
  p_{\tau_0,\Delta\tau}(\tau) :=
  \begin{cases}
    1\qquad\qquad\text{for}\quad |\tau| < \tau_0-\Delta\tau\\
    0\qquad\qquad\text{for}\quad |\tau| > \tau_0+\Delta\tau\\
    \left\{
      1+\exp\left[
        \tan\left(\frac{\pi(|\tau|-\tau_0)}{2\Delta\tau}\right)
        \right]
      \right\}^{-1}
      \text{otherwise}
    \end{cases}
\end{equation}
which was also used to achieve the deconvolutions shown in Fig.\ 2 of Ref.\ 
\onlinecite{raas04a}. The parameter $\tau_0$ determines where the low-pass
cutoff is done; the parameter $\Delta\tau$ determines on which time-scale the
cutoff function switches from $1$ to $0$. We used $\tau_0 \approx 10.7$ and
$\Delta\tau \approx 0.763$. In practice, it turns out that it is fairly obvious
in the $\tau$-domain which values one has to choose for the low-pass filter.
The data for too large values of $|\tau|$ scatter very much.


\subsubsection{Matrix Inversion}

A robust alternative to the deconvolution by Fourier transform is the explicit
matrix inversion of the convolution procedure. This procedure shares the
linearity with the Fourier transform and it may also lead to negative spectral
weight close to abrupt changes of $\rho(\omega)$, for instance at
singularities. An advantage over the Fourier deconvolution is that one may also
consider variable widths $\eta_i$. In principle, this allows to adapt the grid
$\{ \xi_i\}$ to the expected behavior of $\rho(\omega)$. A denser grid can be
taken where the spectral density varies more rapidly.
\begin{figure}[t]
  \begin{center}
    \includegraphics[width=0.98\columnwidth,angle=0]{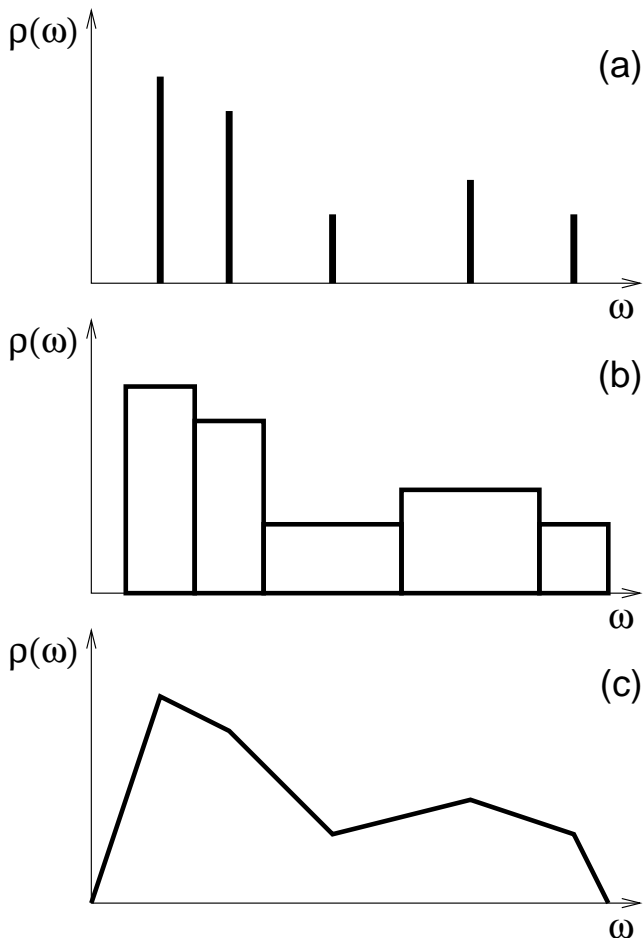}
    \caption{\label{fig:rho} 
      Illustration of various assumptions on the behavior of the spectral
      density $\rho(\omega)$.
      (a) Spikes: $\rho(\omega)$ consists of a set of $\delta$-functions with
      given positions, but unknown weights.
      (b) Histogram: $\rho(\omega)$ is piecewise constant with given positions
      of the jumps, but unknown heights $\rho_i$.
      (c) Continuous: $\rho(\omega)$ is piecewise linear and continuous with
      given positions of the cusps, but unknown values $\rho(\omega_i)$.}
  \end{center}
\end{figure}

The raw data $\{g_i\}$ provided by D-DMRG for a set of $\{\xi_i\}$ are taken as
the components of the vector $\mathbf{g}$. Then the process of convolution can
be described by a linear mapping $\mathbf{M}$ of a set of linear parameters $\{
l_i\}$ characterizing $\rho(\omega)$ onto the raw data $\{ g_i\}$. Let us take
the $\{ l_i\}$ also as components of a vector $\mathbf{l}$. Then the
convolution reads
\begin{equation}
  \label{eq:matrixinvers}
  \mathbf{g}= \mathbf{M}\:\mathbf{l}\ .
\end{equation}
Clearly, the deconvolution implies the inversion of this equation. To this end,
we have to choose the matrix $\mathbf{M}$ to be square which means that there
must be as many data values $g_i$ as there are parameters $l_i$ to be
determined. It is obvious that this analysis is linear. The precise form of the
matrix $\mathbf{M}$ depends on further assumptions. Three generic scenarios are
studied; they are illustrated in Fig.\ \ref{fig:rho}.

\paragraph{Spikes} 
Assuming that $\rho(\omega)$ is given as set of $\delta$-functions (cf.\ Fig.\ 
\ref{fig:rho}a) with weights $\{ w_i\}$ at given frequencies $\{\omega_i\}$.
Then the weights constitute the linear parameters defining the spectral
density. The matrix $\mathbf{M}$ is derived from Eq.\ (\ref{eq:convolv0}); it
has simple matrix elements given by Lorentzians
\begin{equation}
  M_{n,i}=L_{\eta_n}(\xi_n-\omega_i) \ .
\end{equation}
This approach has been proposed and used by Jeckelmann and coworkers
\cite{gebha03,nishi04a}. To have a square matrix problem the number of spikes
has to equal the number of $g_i$ determined by D-DMRG.

The rendering of the results is subtle. Since one cannot plot
$\delta$-functions they have to be broadened which adds another free parameter.
Which broadening one has to choose is not {\it a priori} clear. For equidistant
grids ${\omega_i}$ the distance between two consecutive peaks is a natural
choice \cite{gebha03,nishi04a}, cf.\ Fig.\ \ref{fig:decon1}.

\paragraph{Histogram}
One assumes that $\rho(\omega)$ is piecewise constant $\rho(\omega)=\rho_i$ for
$\omega_i< \omega \le \omega_{i+1}$ (cf.\ Fig.\ \ref{fig:rho}b) where the
frequencies $\omega_i$ are given beforehand. The set of linear parameters
defining the spectral density are the values $l_i=\rho_i$. Again, Eq.\ 
(\ref{eq:matrixinvers}) has to be solved. The matrix elements of $ \mathbf{M}$
are found from the integration of the right hand side of Eq.\ 
(\ref{eq:convolv0})
\begin{equation} 
  M_{n,i}= \pi^{-1} \arctan\left[(\omega-\xi_n)/\eta_n\right]
  \big|_{\omega=\omega_i}^{\omega_{i+1}}\ .
\end{equation}
The number of values $\rho_i$ (not the number of $\omega_i$) must be equal to
the number of $\xi_i$ to ensure that $ \mathbf{M}$ is square. The rendering
is straightforward since the values $\rho_i$ represent densities which can be
plotted directly, cf.\ the histograms in Fig.\ \ref{fig:decon1}.

\paragraph{Continuous}
One assumes that $\rho(\omega)$ is continuous and piecewise linear (cf.\ Fig.\ 
\ref{fig:rho}c) with $\rho(\omega_j)=\rho_j, j\in\{1,2,3,\ldots, p\}$ at given
frequencies $\omega_i$. The values $\rho_i$ represent the linear parameters
$l_i=\rho_i$ which are determined by Eq.\ (\ref{eq:matrixinvers}). The matrix
elements result from the integration in Eq.\ (\ref{eq:convolv0}). They are
given by $M_{n,i}=\partial_{\rho_i} A_n(\xi_n)$ with
\begin{subequations}
  \begin{align}
    \begin{split}
      A_n(\xi) &= \sum_{j=1}^{p-1} 
      \left\{\eta_n a_j \ln[\eta_n^2+(\xi-\omega)^2]/(2\pi)+\right.\\
      &\phantom{=}\left.(\xi a_j+b_j) \arctan[(\omega-\xi)/\eta_n]/\pi
      \right\}
      \big|_{\omega=\omega_j}^{\omega_{j+1}}
    \end{split}\\
    a_j &= (\rho_{j+1}-\rho_j)/(\omega_{j+1}-\omega_j)\\
    b_j &= (\omega_{j+1}\rho_j-\omega_j\rho_{j+1})/(\omega_{j+1}-\omega_j)\ .
  \end{align}
\end{subequations}
The number $p$ must be equal to the number of raw data points. An example is
depicted by the piecewise linear dashed curve in Fig.\ \ref{fig:decon1}.

In all schemes, the numbers must be chosen such that $\mathbf{M}$ is a square
matrix. This is a necessary but not a sufficient condition for the existence of
a unique solution $\mathbf{l}$ in Eq.\ (\ref{eq:matrixinvers}). In practice, we
did not encounter problems in the inversion of Eq.\ (\ref{eq:matrixinvers}) as
long as the raw data points were distributed rather evenly along the real axis.
Only if there are data points accumulating in certain regions, for instance
several data points at the same frequency, the inversion can be problematic.
Loosely speaking this may occur since the raw data is slightly contradictory
due to numerical inaccuracies. The broadening by the $\{\eta_n\}$ reduces the
differences between the spectral densities. Hence small deviations in the raw
data have large effects on the extracted spectral densities.

All linear extraction schemes do not guarantee that the extracted spectral
density is non-negative, see Fig.\ \ref{fig:decon1}. Whether this must be
considered a serious drawback depends on the extent to which negative values
occur and on the context in which the result is used. If the spectral density
is the final result small regions of overshooting are unproblematic. If,
however, the overshooting is considerable and if the spectral density shall be
used in a subsequent step, e.g.\ in the self-consistency of a DMFT calculation,
then negative values pose a severe problem. An important example is the
determination of the coefficients of the continued fraction of the spectral
density. This determination is only possible if the spectral density is really
non-negative.

Another problem is the smoothness of the extracted density. Spurious
discontinuities like the ones assumed in the spike ansatz or in the histogram
ansatz (see Figs.\ \ref{fig:decon1} and \ref{fig:rho}) imply singularities in
the real part of the propagator which is determined by the Kramers-Kronig
relation. This in turn leads to unwanted features like slowly decaying
oscillations in coefficients of the continued fraction. Of course, various
schemes can be used to interpolate the discrete data provided by the matrix
inversion approaches. But the interpolation represents an additional
approximation which can be difficult to control.


\subsection{Nonlinear Extraction Schemes}


\subsubsection{Basic Algorithm}

In view of the drawbacks of the linear extraction schemes it is worthwhile to
think about alternatives. The objective is to devise an ansatz for a
continuous, non-negative spectral density $\rho(\omega)$ which is consistent
with the numerically determined values of the raw data $\{ g_i\}$ in Eq.\ 
(\ref{eq:convolv0}). The ideal ansatz is completely unbiased. That means it
does not use \emph{any} information \emph{other} than the one provided by the
raw data. The information content of a density $\rho(\omega)$ is measured up to
a constant by its negative entropy
\begin{equation}
  \label{eq:entropy}
  -S = \int_{-\infty}^\infty\rho(\omega)\ln(\rho(\omega))\text{d}\omega\ .
\end{equation}
The least biased ansatz is the one with the least information content which is
still compatible with the raw data. Hence we have to look for the density
$\rho(\omega)$ which minimizes $-S$ (maximizes $S$) under the conditions
(\ref{eq:convolv0}) given by the raw data $\{ g_i\}$ and by the known
normalization
\begin{equation}
  \label{eq:norm}
  1 = \int_{-\infty}^\infty\rho(\omega)\text{d}\omega\ .
\end{equation}

To find this least biased ansatz (LB) is a straightforward task. Using the
Lagrange multipliers $\{\lambda_i\}$ for the $p$ conditions set by the raw data
$\{ g_i\}$ and the Lagrange multiplier $\widetilde\mu$ for the normalization
(\ref{eq:norm}) the least biased ansatz is characterized by $\delta S =0 $,
i.e.\ 
\begin{equation}
  0=-1-\ln(\rho(\omega)) +\sum_{i=1}^p \lambda_i L_{\eta_i}(\omega-\xi_i)
  +\widetilde\mu \ .
\end{equation}
This equation implies that the LB ansatz reads
\begin{equation}
  \label{eq:LB}
  \rho(\omega) = \exp\left[
    \mu+\sum_{i=1}^p \lambda_i L_{\eta_i}(\omega-\xi_i)
  \right]\ ,
\end{equation}
where we defined $\mu=\widetilde\mu-1$. The Lagrange multipliers are determined
by the nonlinear equations (\ref{eq:norm}) and (\ref{eq:convolv0}). They can be
determined by any standard algorithm for a set of nonlinear equations. Via the
ansatz (\ref{eq:LB}), the $p+1$ Lagrange multipliers determine the most
unbiased spectral density $\rho(\omega)$ which is still compatible with the
numerically measured information on $\rho(\omega)$.

The LB ansatz (\ref{eq:LB}) is positive and continuous. Hence it avoids two
major drawbacks of the linear extraction schemes. In spite of its continuity
the LB ansatz is governed by a restricted number of parameters. Despite its
positivity, the LB ansatz is able to reproduce rather abrupt changes in the
spectral density, see Fig.\ \ref{fig:decon1}. If arbitrarily accurate data at
a fixed value of $\eta$ were available on an arbitrarily dense grid, the LB
algorithm were able to provide the correct result with arbitrary resolution
(see also the discussion of Fig.\ \ref{fig:decon3} below). But the accuracy of
the raw data required to achieve a certain resolution in the deconvolved result
grows exponentially. So in practice this route cannot be followed very far and
the broadening $\eta$ sets the scale for the achievable resolution.

The least bias approach belongs to the class of maximum entropy methods
(MaxEnt) \cite{press92}. The main difference to standard MaxEnt is that we do
not use a $\chi$-functional in addition to the entropy function
(\ref{eq:entropy}). The $\chi$-functionals are bilinear in the density. They
are introduced to account for possible deviations of the $g_i$ from their true
values. Such deviations occur for instance in quantum Monte Carlo calculations
due to the inevitable statistical error. The D-DMRG data is free from
statistical errors. Hence we can use the entropy functional alone as described
above. The correction of numerical inaccuracies is discussed in more detail
below.

We emphasize a major difference between the extraction of the spectral density
$\rho(\omega)$ from D-DMRG data and from QMC. In the former case the task to be
solved is to remove a small imaginary part $\omega+i\eta \to \omega$. This
means that in the D-DMRG the raw data is situated slightly above the real axis
and has to be continued down to it. In the latter case the task to be solved is
to continue QMC data from the imaginary axis (no real part) to the real axis
$i\zeta \to \omega$. So the challenge in the QMC data analysis is much greater
than in the D-DMRG data analysis. This explains why the D-DMRG approach is
suited to investigate sharp features also at high energies \cite{raas04a} while
this is not a straightforward task by QMC.

The solution of the nonlinear equations (\ref{eq:norm}) and (\ref{eq:convolv0})
is done by a standard algorithm. There is no mathematical argument to show that
there is only one unique solution. For instance, we found that the
normalization of the spectral density need not be ensured by the parameter
$\mu$. This parameter can be fixed to almost any value. The remaining Lagrange
parameters suffice to determine good approximations to $\rho(\omega)$ which
fulfill the normalization condition (\ref{eq:norm}) well. This implies that
iterative numerical solutions have difficulties to fix $\mu$ independently. The
root-finder algorithms run much more stable if the normalization is not
included in the set of equations. Yet the resulting densities are normalized
if the raw data provides a reasonable scan of the spectral density, i.e.\ if
there is raw data at all frequencies $\omega$ where the density is
non-negligible.

For other sum rules, for instance the second moment $\langle \omega^2\rangle$,
the same conclusion holds. If the raw data scans all relevant frequencies
the sum rules do not provide useful additional information.
But if raw data is only available for restricted frequency intervals, the
sum rules help to improve the LB ansatz.

Furthermore, the iterative numerical solutions depend sometimes on the initial
values. But the resulting $\rho(\omega)$ are in general (almost) identical.
This remains true if the iterative algorithm does not find a true solution of
the set of non-linear equations but only a set of parameters which makes the
deviations
\begin{equation}
  \label{eq:difference}
  \Delta g_i := g_i - \frac{1}{\pi}\
  \int_{-\infty}^\infty 
  \frac{\eta_i \rho(\omega)\text{d}\omega}{(\xi_i-\omega)^2+\eta_i^2}
\end{equation}
small but fails to make them zero. In summary, the iterative determination of
the Lagrange multipliers does not represent a major problem.

In Fig.\ \ref{fig:decon1} we display a comparison of all data extraction
schemes introduced so far. The line shape to be found is a singular power law
divergence. This line shape constitutes an unsolvable task since a divergence
cannot be reproduced by the algorithms discussed. But this example illustrates
well to which extent the algorithms manage to render the true distribution of
spectral weight. All the linear schemes lead to regions of negative spectral
densities which is a severe drawback. The divergence is approximated by a
broad peak located at $0.05$ away from the position of the divergence. There
are some spurious oscillations in the approximated spectral density.

The LB scheme avoids negative spectral weight by construction. The divergence
is approximated by a sharper peak at about $0.02$ away from the position of the
divergence. The algorithm implies spurious oscillations in the approximated
spectral density. So we conclude that the LB analysis represents a very
efficient reconstruction of the unbroadened data even though the spurious
oscillations can be a nuisance.

On the other hand it is, of course, possible to improve the analysis. An
additional data point at $\omega\approx 0.02$ will certainly help all
algorithms to reproduce the unbroadened density more faithfully. The same is
trivially true for a denser mesh of raw data points. The former solution,
however, requires either to intervene manually in the data analysis or to know
beforehand where the peaks will be located. The latter requires much more
numerical effort on the D-DMRG level so that this is not the most efficient
approach.

\begin{figure}[t]
  \begin{center}
    \includegraphics[width=\columnwidth,angle=0]{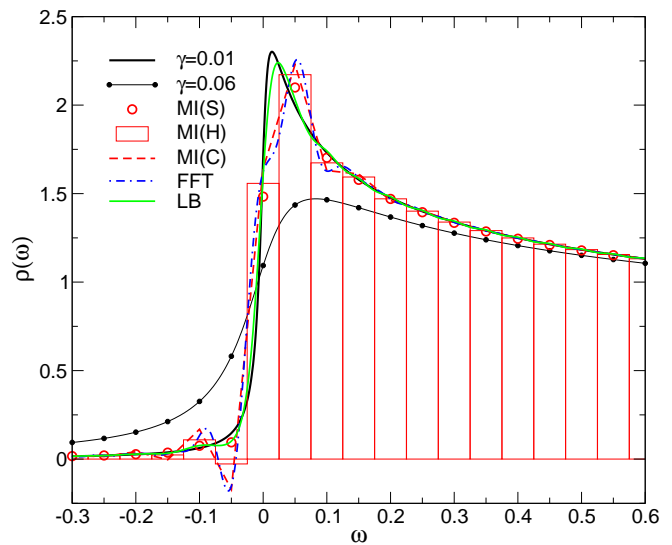}
    \caption{\label{fig:decon2} 
      Exact (thick black solid line) and broadened (by $\eta=0.05$, thin black
      solid line with circles representing the raw data at $\xi_i=0.05 i$) 
      lines
      derived from the Doniach-\v{S}unji\'c line shape in Eq.\ 
      (\ref{eq:doniach}), see also main text. The results of various schemes to
      retrieve the unbroadened line are shown: by deconvolution via fast
      Fourier transform (FFT), by matrix inversion assuming spikes (MI(S)) or
      histograms (MI(H)) or a piecewise linear continuous function (MI(C)), by
      the non-linear least-bias algorithm (LB).}
  \end{center}
\end{figure}
In Fig.\ \ref{fig:decon2} we present the analysis of a smooth exact curve,
namely the line shape in (\ref{eq:doniach}) at $\gamma=0.01$. The curve
broadened by $\eta=0.05$ is the one at $\gamma=0.06$ since the widths of
Lorentzians is additive under convolution, in contrast to the root-mean-square
of narrower distribution functions. Clearly, the extraction schemes do a better
job for this non-singular case. The regions of negative spectral density in the
linear schemes has shrunk. The LB scheme manages to reproduce the true density
almost perfectly. The spurious oscillations are negligible. If we had chosen
$\gamma=0.02$ for the exact curve the LB density would be hardly
distinguishable from the exact curve.

\subsubsection{Robustness towards Inaccuracies}
\label{sec:robust}

\begin{figure}[t]
  \begin{center}
    \includegraphics[width=\columnwidth,angle=0]
    {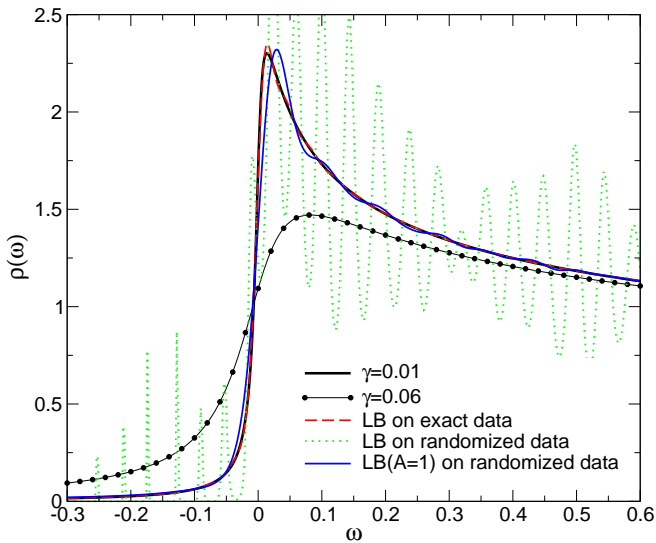}
    \caption{\label{fig:decon3} 
      Exact (thick black solid line) and broadened (by $\eta=0.05$, thin black
      solid line with circles representing the raw data at $\xi_i=0.02i$) lines
      derived from the Doniach-\v{S}unji\'c line shape in Eq.\ 
      (\ref{eq:doniach}). The dashed line depicts the density extracted from
      exact raw data by the LB scheme (the curve lies on top of the thick black
      line except at the maximum). The dotted oscillations are the LB result
      from data contaminated by a random error, see main text. The thin solid
      line represents the density derived from the contaminated data by the
      robust LB scheme with $A=1$ using Eqs.\ 
      (\ref{eq:robustLB},\ref{eq:selconrobust}).}
  \end{center}
\end{figure}
So far we analyzed ideal raw data, i.e.\ no errors were considered.
Statistical errors do not occur in the D-DMRG approach but systematic errors
occur. There are two main sources for such errors. The first is the inaccuracy
of the algorithm due to the truncation of the Hilbert space. This is an
unavoidable error; but it can be controlled by comparing the results for
different numbers of states kept in the density-matrix renormalization. We
perform our calculations for 128 and for 256 states using the representation of
spinful fermions as two kinds of spins via a double Jordan-Wigner
transformation \cite{raas04a}. The relative truncation error in the spectral
densities is estimated to be of the order of $10^{-5}$ to $10^{-3}$ depending
on the frequency where it is computed. For low values of the frequency
($|\omega| \lessapprox D$ in our model) the lower error applies; for
frequencies beyond $D$ the larger value applies.

The second important source of inaccuracy are finite-size effects. In
principle, we wish to compute the spectral density for the infinite system.
But this is not feasible numerically. So the system -- the infinite chain -- is
approximated by finite chains of $L=120$ to 400 sites. In a rigorous sense, the
spectral density of the finite system consists of $\delta$-functions, i.e.\ it
is not a continuous function. But we do not intend to resolve all the fine
details of the finite system. Rather we interpret our numerical raw data
obtained for the finite chain as an approximate description of the infinite
system. The deviation of the raw data for the finite chain from the desired
raw data of the infinite chain is considered the source of a systematic error,
the finite-size effect. There are two conceivable ways to deal with this
error. One way is to perform an extrapolation in system size $L$ for each raw
data point $g_i$ before the deconvolution is done. The other way is to
deconvolve the raw data for various chain lengths and to check whether the
results still depend on the length $L$ .

In our work, we have chosen the second approach because the first is hampered
by an unsystematic behavior of $g_i$ on $L$. Depending on details a particular
$\delta$-peak of the finite systems contributes more or less to the $g_i$ under
study. This makes a controlled extrapolation for all $\{ g_i\}$ difficult, if
not impossible.

In the second approach, care must be taken that the length $L$ is so large that
the raw data is sufficiently close to the raw data of the infinite system. In
practice, this puts a restriction on $\eta$ and $L$, see Sect.\ \ref{sec:kondo}
and Ref.\ \onlinecite{jecke02}. Of course, the use of finite chains restricts
the extent to which we may extract information on the exact infinite system from
the broadened data obtained for the finite system.

Another sort of errors are rounding errors. But they are of minor importance
compared to the two other sources discussed above.

In Fig.\ \ref{fig:decon3} we display the LB analysis of raw data for the exact
curve at $\gamma=0.01$. In contrast to the procedure in Fig.\ \ref{fig:decon2}
we use a finer grid of $\xi_i = i*0.02$ for the raw data (curve with circles).
The extracted curve represents the exact one very well, see the dashed line in
Fig.\ \ref{fig:decon3}. The agreement is significantly better than the one
reached in Fig.\ \ref{fig:decon2}. This illustrates that sufficiently accurate
broadened data at fixed $\eta$ can be used to resolve features of widths below
$\eta$.

As explained above, there are in practice restrictions to better resolutions
due to the systematic errors, namely the truncation of the basis and the
finite-size effect. To examine the effect of systematic errors on the LB
deconvolution we deliberately contaminated the raw data by an error of the
order of $10^{-3}$ according to
\begin{equation}
g_i \to g_i*(1+10^{-3}x)
\end{equation}
where $x$ is a random number between $-1$ and $1$. The randomness is just used
to mimic a systematic error which is uncorrelated from frequency to frequency.
We obtained qualitatively very much the same results for a non-random error
$g_i \to g_i*(1+10^{-3}\cos(\sqrt{5}i))$.

If the data is slightly inaccurate the deconvolution indeed fails as
illustrated by the wild oscillations of the dotted line. We conclude that the
occurrence of strong oscillations can be taken as criterion that the used raw
data is not accurate enough for the LB analysis, i.e.\ the systematic errors
are too large.

The thinner solid line depicts a successful deconvolution of the contaminated
raw data. It is achieved by a modification of the LB algorithm which makes it a
standard maximum entropy approach. The negative entropy functional
(\ref{eq:entropy}) is supplemented as shown
\begin{align}
  \label{eq:robustLB}
  F[\rho(\omega)] :&= -S[\rho(\omega)] + A \chi[\rho(\omega)]\\
  &= \int_{-\infty}^\infty \rho(\omega) \ln(\rho(\omega)) \text{d}\omega
  + A\sum_i (\Delta g_i)^2
  \nonumber
\end{align}
by a quadratic functional $\chi$ which measures the distance from the perfect
fulfillment of the conditions (\ref{eq:convolv0}). The differences $\Delta
g_i$ are defined in Eq.\ (\ref{eq:difference}). The minimization of the
functional $F$ leads to the same ansatz (\ref{eq:LB}) as before except that the
parameters $\lambda_i$ are now given by
\begin{equation}
\label{eq:selconrobust}
\lambda_i = 2A\Delta g_i \ . 
\end{equation}
The set of these nonlinear equations (instead of $\Delta g_i=0$ as for the pure
LB approach) is used to determine the parameters $\lambda_i$. It is obvious
that the robust modification of the LB ansatz becomes the pure LB ansatz in the
limit of $A\to\infty$ since in this limit the deviations $\Delta g_i$ vanish
for given values of the Lagrange parameters $\lambda_i$. If the data is only
weakly contaminated by inaccuracies, large values of $A$ can be used to extract
the spectral densities. Our example in Fig.\ \ref{fig:decon3} shows fairly
strongly perturbed data. Still the robust LB can make sense out of them and
retrieves a good approximation to the underlying curve. 

The modified LB ansatz (\ref{eq:robustLB}) is more robust since it can deal
with some inaccuracies or inconsistencies of the raw data. Imagine that we
deal with raw data on a fine grid where the distance between the data points is
significantly smaller than the Lorentzian width $\eta_i \gg (\xi_{i+1}-\xi_i)$. 
Then $g_{i+1}$ and $g_{i}$ may differ only slightly if they are derived from a
smooth continuous density $\rho(\omega)$. Any inaccuracy spoils this relation
and may introduce significant spurious oscillations, see Fig.\ 
\ref{fig:decon3}. As we stressed already previously the broadening makes
different data more alike. Hence, the inverse process enhances slight
differences like the ones between exact and inaccurate raw data greatly. The
robust LB ansatz (\ref{eq:robustLB}) helps to make the data extraction less
sensitive to such effects without losing much resolution. Thereby, spurious
oscillations can be suppressed. 

The robust LB scheme opens the possibility to resolve features of widths below
a given value of the broadening $\eta$ by using a finer grid with $\Delta \xi <
\eta$ since slightly inaccurate data can still be deconvolved. In this way, one
may avoid the explicit use of small values of $\eta$. We emphasize, however,
that the data must be accurate enough to contain the information on the
relevant physics. Of course, the robust LB approach is no means to extract
information which is not given by the raw data. For instance, one may stick to
short chain lengths \emph{only} if the underlying physical problem does not
demand to describe long-range spatial fluctuations. 


\section{Kondo Resonance}
\label{sec:kondo}
Next we present data for the local spectral density $\rho(\omega)$ for the SIAM
as introduced in Sect.\ \ref{sec:model}. In the present section we focus on the
behavior at low energies where $\rho(\omega)$ is dominated by the Kondo peak,
see e.g.\ Ref.\ \onlinecite{hewso93}. This peak can be computed by NRG in a
very efficient way \cite{sakai89,costi90,hewso93}. So the idea of the present
calculation is not to provide novel data, but to gauge the D-DMRG and to
demonstrate that features at low energies can be resolved. We use the LB scheme
to extract the real spectral density from the raw data as explained above. This
is one of the differences to the investigation by Nishimoto and Jeckelmann
\cite{nishi04a}. For details of the numerical realizations we refer to previous
works \cite{raas04a,nishi04a}. Another difference concerns the parameter
regime. We consider here interaction values $U$ beyond the bare band width
$W=2D$, i.e.\ $U>2D$, while Nishimoto and Jeckelmann look at Lorentzian bare
spectral function whose band width is very large. But for the low energy region
this difference is only a quantitative one. 

\begin{figure}[t]
  \begin{center}
    \includegraphics[width=\columnwidth,angle=0]{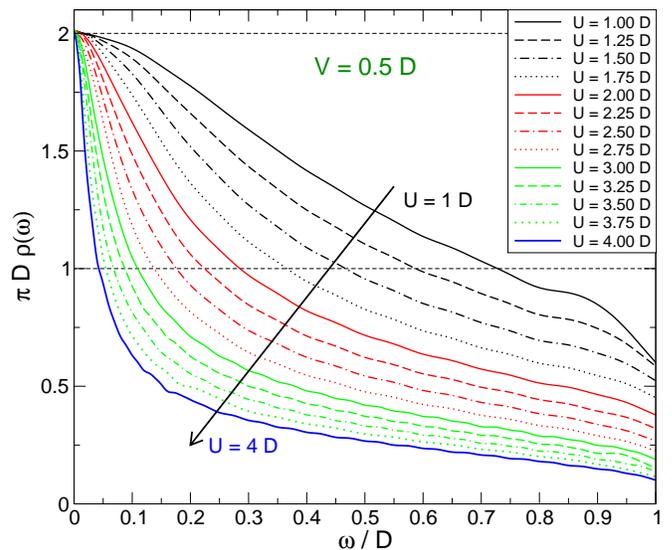}
    \caption{\label{fig:kondo} 
      Kondo peaks for the single-impurity Anderson model (\ref{eq:hamilton}) 
      for a hybridization function with semi-elliptic spectral density and
      $V=D/2$. On increasing interaction the Kondo peak becomes rapidly 
      narrower.} 
  \end{center}
\end{figure}
Our results are depicted in Fig.\ \ref{fig:kondo}. We have chosen the
parameters such that the hybridization function has a semi-elliptic spectral
density. The hybridization is taken to be $V=D/2$ so that the spectral density
of the $d$-site without interaction is semi-elliptic, too. Clearly visible is
the rapid narrowing of the Kondo peak on increasing interaction. Note that the
density at zero energy $\rho(0)$ is pinned to its non-interacting value as
required by the sum rules \cite{lutti60b,lutti61,ander91,hewso93}. This exact
result is fulfilled to numerical accuracy which ranges from $0.1\%$ for smaller
interactions to $1\%$ at larger interactions, see Fig.\ \ref{fig:kondo}. 

The data in Fig.\ \ref{fig:kondo} is obtained on various grids in frequency,
for various chain lengths and for varying widths $\eta$. All calculations are
done with $m=128$ sites kept in the truncations of the DMRG. The chain lengths
$L$ vary between 120 and 400 fermionic sites which translates to 240 (e.g.\ at
$U=1.25D$) to 800 (e.g.\ at $U=4.00D$) spin sites after the double
Jordan-Wigner transformation. The smaller the widths $\eta$ are chosen the
longer the chains have to be taken \cite{hovel00,jecke02}. A chain of $L$
fermionic sites implies $L$ main peaks distributed over the band width $W=2D$. 
Assuming a roughly equidistant distribution the distance between two main peaks
is $W/L$ which should be significantly smaller than $\eta$
\begin{equation}
  2W/L \le \eta
\end{equation}
in order to ensure that the discrete structure of the finite system is
sufficiently smeared out. Then the data provided can be interpreted reasonably
well as data of the infinite system. 

Since the calculations are less costly at low energies than at higher energies
we used mixed raw data coming from various chain lengths. The width $\eta$ is
varied correspondingly; we always used $\Delta \xi =\xi_{i+1}-\xi_i = \eta$. 
For instance at $U=4D$, we used $L=400$ with $\eta=0.01D$ between $\omega=0.00$
and $0.05$; $L=200$ with $\eta=0.02D$ between $\omega=0.06$ and $0.18$; $L=200$
with $\eta=0.05D$ between $\omega=0.20$ and $2.95$; $L=120$ with $\eta=0.10D$
between $\omega=3.00$ and $4.00$. The analysis of the raw data was done in all
cases by the pure LB ansatz (\ref{eq:LB}) with the conditions $\Delta g_i=0$. 
There was no need to use the robust LB with $\lambda_i=2A\Delta g_i$. 

The rapidly narrowing peaks in Fig.\ \ref{fig:kondo} are characterized by the
Kondo energy scale $T_\text{K}$. This scale can be read off from the spectral
densities, for instance as half the width at half the maximum, i.e.\ at $\pi
D\rho(\omega=T_\text{K})=1$. From analytic considerations \cite{hewso93}, it is
known that the Kondo energy scale is exponentially small in the interaction
\begin{equation}
  \label{eq:kondoscale}
  T_\text{K} \propto V \sqrt{D/U}\exp\left(-\pi UD/(4V)^2\right)\ . 
\end{equation}
This formula applies in the limit of $U \gtrapprox W=2D$ which is called the
Schrieffer-Wolff limit of the SIAM. It holds for $U\to\infty$. 
\begin{figure}[t]
  \begin{center}
    \includegraphics[width=\columnwidth,angle=0]{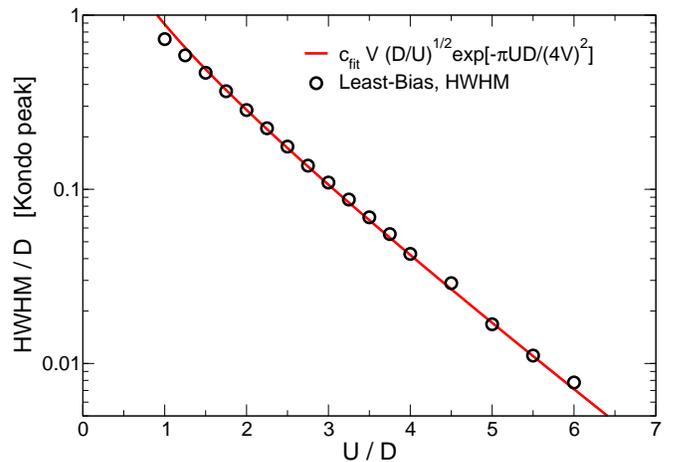}
    \caption{\label{fig:kondoscale} 
      Kondo energy scale (Kondo temperature) for the single-impurity Anderson
      model (\ref{eq:hamilton}) for a hybridization function with semi-elliptic
      spectral density and $V=D/2$ in the Schrieffer-Wolff limit $U>2D$. 
      Symbols: half-width-half-maximum (HWHM) read off from $\rho(\omega)$ as
      found by the LB analysis of D-DMRG raw data; line: analytic result in
      Eq.\ \ref{eq:kondoscale}. The proportionality factor of the fit is
      $c_\text{fit}=3.887$.} 
  \end{center}
\end{figure}
The numerical results for the Kondo scale and the analytical prediction
(\ref{eq:kondoscale}) agree very well in Fig.\ \ref{fig:kondoscale}. Only for
$U<W$ deviations occur as was to be expected. The widths at large values of $U$
($U>4D$) are found from results for $\rho(\omega)$ where we extracted only the
behavior at low energies. To determine the HWHM it is not necessary to know the
whole line shape. It is an asset of the LB extraction scheme that it allows
also to determine only a part of the whole curve. In conclusion, Fig.\ 
\ref{fig:kondoscale} demonstrates that the dynamic DMRG is able to reproduce
the low energy scale of the SIAM over two orders of magnitude. 


\section{Hubbard Satellites}
\label{sec:satellit}
In this section we show that the D-DMRG combined with the powerful LB scheme
allows to resolve features at high energies which so far eluded a quantitative
determination. The calculation of the line shapes of energy levels at high
energies represents a new field of applications since previous methods are not
suited to perform such computations. An exception are features at high energies
which can be understood and described as shifted low energy features
\cite{helme05}. We emphasize that the progress in the manipulation of quantum
dots by optical means has brought the measurements of high energy features
within reach, see e.g.\ Refs.\ \onlinecite{warbu00} and \onlinecite{karra04}. 

\begin{figure}[t]
  \begin{center}
    \includegraphics[width=\columnwidth,angle=0]{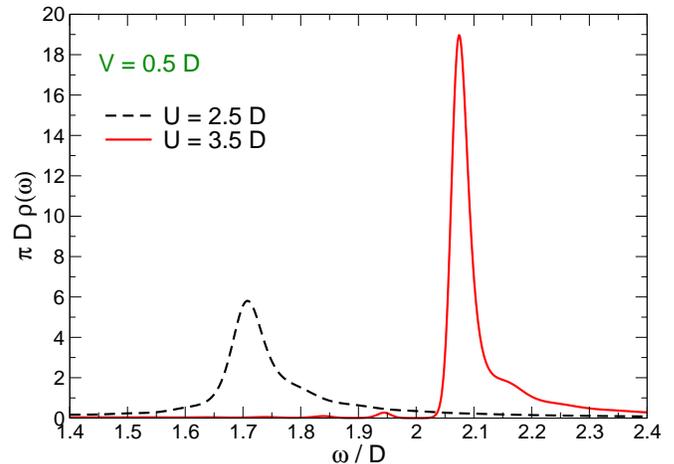}
    \caption{\label{fig:satellit} 
      Upper Hubbard satellite for two values of the interaction $U$
      obtained from D-DMRG raw data ($\eta=0.1D,\Delta\xi= 0.05D, 
      m=256, L= 120$)
      processed by the LB scheme. 
      At larger $U$ a strongly asymmetric line shape occurs.} 
  \end{center}
\end{figure}
In a previous work \cite{raas04a}, we have shown that the Hubbard satellites at
about $\omega=\pm U/2$ become increasingly sharp once the interaction $U$ is
larger than the bandwidth $W=2D$. In that work, however, we were not able to
resolve the sharp Hubbard satellites since the deconvolution by Fourier
transform was not powerful enough to do so, cf.\ Sect.\ \ref{sec:extract}. By
means of the LB extraction we are now able to address the line shape. In Fig.\ 
\ref{fig:satellit} the results for a moderate value of $U$ ($2.5D$) and for a
larger value of $U$ ($3.5D$) are shown. While the line at $U=2.5D$ is still
fairly symmetric (cf.\ also the line for $U=2D$ in Fig.\ 2 of Ref.\ 
\onlinecite{raas04a}) the line at $U=3.5D$ has a clear asymmetric shape. The
rise at the low energy side is rather abrupt and steep while the decrease at
the high energy side is much more gentle and slow. The peak is very pronounced
and the maximum value very high. The small bumps left of the peak are artefacts
of the data extraction similar to the small dip in front of the uprise in Fig.\ 
\ref{fig:decon2}. We also consider the shoulder on the right side of the peak
in Fig.\ \ref{fig:satellit} to be an artefact of the data analysis in analogy
to the spurious oscillations in Fig.\ \ref{fig:decon1}. 

At present, we do not have an analytical description of the line shape in Fig.\ 
\ref{fig:satellit}. Qualitatively, however, the following description holds. 
The doubly occupied level has the energy $U/2$ in the atomic limit for $V=0$. 
This level is shifted to higher energies by a finite hybridization $V>0$
\cite{schri66,raas04a} and it is broadened \cite{raas04a}. The shift $\propto
V^2/U$ is due to level repulsion between the doubly occupied level and modes
where the doubly occupied level excites additional particle-hole pairs. A
rough, averaged description of the broadening is provided by Fermi's golden
rule implying that the width is proportional to $V^4/U^2$. 

In view of the asymmetric line shape and of the very narrow peak this averaged
picture can be improved. The narrow peak itself has a certain intrinsic width
which is small. Note that the general understanding of the SIAM as a local
Fermi liquid implies that there are no singularities in the propagator. The
spectral density is a smooth function, though it may display very narrow
features. This is in contrast to the X-ray edge problem where the added fermion
is completely immobile (no recoil). 

The intrinsic width of the narrow peak results from the decay of the doubly
occupied level into particle-hole pairs of the local Fermi liquid. According to
Fermi's golden rule such a decay can take place only if the energy of the
initial and the final state is equal which means that many particle-hole pairs
have to be excited in order to make up for the relatively large energy $\approx
U/2$ of the double occupancy. Very many particle-hole pairs are needed in view
of the reduced effective band width of the order of the Kondo energy scale
$T_\text{K}$. This argument accounts for a finite, though very large, life
time of the doubly occupied level as seen in Fig.\ \ref{fig:satellit}. 

The high energy tail of the line in Fig.\ \ref{fig:satellit} can be understood
as an effect of particle-hole pairs which are generated by the double
occupancy. This is \emph{not} the same as the decay of the double occupancy
described above. In the decay the double occupancy disappears in the process
while it remains on generating additional particle-hole pairs. This explains
why the contribution of this process is found at the high energy side of the
peak in Fig.\ \ref{fig:satellit}. The additional particle-hole pairs require an
additional amount of energy to be created. The physics of this process is
similar to the physics of the X-ray edge problem where the change of a local
potential induces infinitely many particle-hole pairs thus leading to slowly
decreasing tails in the spectra, see for instance Ref.\ \onlinecite{schot69b}
and references therein. The main difference to the X-ray edge problem is that
there is no true singularity here so that the line shape is smoother. 

At present, the accuracy of the data in the high energy tails is not sufficient
to search for approximate power laws which possibly describe the tails of the
Hubbard satellites in Fig.\ \ref{fig:satellit}. 


\section{Summary}
\label{sec:summary}

In this article, we discussed a variety of schemes to extract the spectral
density $\rho(\omega)$ from the results of dynamic density-matrix
renormalization data. All these schemes have the aim to remove the unavoidable
broadening which has to be included in a D-DMRG calculation. The linear schemes
use either Fourier transform to deconvolve the raw data or they implement an
explicit matrix inversion. These schemes are linear because there is a linear
relationship between the raw data and the extracted spectral density. If the
structures to be resolved are not too sharp the linear schemes work well. If
there are sharp structures the linear schemes are prone to lead to negative
spectral densities which result from spurious oscillations. Furthermore, they
can resolve the positions of sharp peaks only with the accuracy of the grid on
which the raw data has been computed. 

The nonlinear scheme introduced belongs to the family of maximum entropy
methods. If the raw data is sufficiently accurate the least-bias approach works
very well. It provides a positive and continuous ansatz for the spectral
density with the least possible bias. Even relative abrupt changes of the
spectral density can be reproduced satisfactorily. In the vicinity of
singularities spurious oscillations occur. But they do not violate the
positivity of the ansatz. The least-bias ansatz can be made more robust towards
small numerical inaccuracies and finite-size effects by including besides the
entropy functional a $\chi$-functional in the functional to be minimized. 
Thereby, one can allow for small deviations from the raw data. 

The properties of the above mentioned schemes were illustrated by calculations
for a singular toy spectral density of the Doniach-\v{S}unji\'c type. D-DMRG
calculations were carried out and presented for the Kondo energy scale of the
symmetric single impurity Anderson model in the Schrieffer-Wolff limit of $U>W$
where $U$ is the interaction and $W$ the band width. The Kondo energy scale
could be retrieved over two orders of magnitude. 

The reproduction of the Kondo energy scale served as a gauge for the approach. 
To our knowledge, numerically exact results like those obtained for the line
shape of the Hubbard satellites at $\omega = \pm U/2$ beyond the region of the
bare band $\omega\in[-D,D]$ are not available in the literature. By these line
shapes we extended our previous analysis of the Hubbard
satellites\cite{raas04a}. Furthermore, these line shapes illustrate that the
dynamic density-matrix renormalization is suited to provide high resolution
data at high energies, i.e.\ away from the Fermi level. It is to be expected
that such information will become more and more important as the experimental
techniques are improving. 

\bigskip


\begin{acknowledgments}
  We thank T.~Costi, R.~Helmes, F.~Gebhard, Y.~Kuramoto, E.~M\"uller-Hartmann,
  A.~Rosch and L.~H.~Tjeng for helpful discussions and the DFG for financial
  support in SFB 608. 
\end{acknowledgments}


\end{document}